\documentclass[
]{ceurart}

\usepackage{float}
\usepackage{listings}
\usepackage{amssymb}
\usepackage{todonotes}
\usepackage{textcase}

\definecolor{specializedcliniccolor}{RGB}{0, 153, 0}
\definecolor{pharmacolor}{RGB}{0, 110, 175}% Define a custom color (in this case, red)
\definecolor{scriptcolor}{RGB}{237, 237, 237}
\newcommand{\Actor}[1]{\textrm{\MakeTextLowercase{#1}}}
\newcommand{\Compo}[1]{\texttt{#1}}

\usepackage{comment}

\usepackage[T1]{fontenc}
\usepackage{lmodern}

\usepackage{orcidlink}
\usepackage[capitalise,nameinlink]{cleveref}
\crefformat{enumi}{#2\textup{#1}#3}

\usepackage{paralist}

\definecolor{specializedcliniccolor}{RGB}{0, 153, 0}
\definecolor{pharmacolor}{RGB}{0, 110, 175}% 

\begin{document}

\copyrightyear{2024}
\copyrightclause{Copyright for this paper by its authors.
  Use permitted under Creative Commons License Attribution 4.0
  International (CC BY 4.0).}

\conference{ICPM 2024 Tool Demonstration Track, October 14-18, 2024, Kongens Lyngby, Denmark}

\title{CONFINE: Preserving Data Secrecy in Decentralized Process Mining}

%\tnotetext[1]{You can use this document as the template for preparing your publication.}

\author[1]{Valerio Goretti}[orcid=0000-0001-9714-4278]
\author[1]{Davide Basile}[orcid=0000-0002-5804-4036]
\author[1]{Luca Barbaro}[orcid=0000-0002-2975-5330]
\author[2]{Claudio Di Ciccio}[orcid=0000-0001-5570-0475]

\address[1]{Sapienza Univeristy of Rome, Italy}
\address[2]{Utrecht University, Netherlands}

\cortext[1]{Corresponding author.}
\fntext[1]{These authors contributed equally.}

\begin{abstract}
In the contemporary business landscape, collaboration across multiple organizations offers a multitude of opportunities, including reduced operational costs, enhanced performance, and accelerated technological advancement. The application of process mining techniques in an inter-organizational setting, exploiting the recorded process event data, enables the coordination of joint effort and allows for a deeper understanding of the business. Nevertheless, considerable concerns pertaining to data confidentiality emerge, as organizations frequently demonstrate a reluctance to expose sensitive data demanded for process mining, due to concerns related to privacy and security risks. The presence of conflicting interests among the parties involved can impede the practice of open data sharing. To address these challenges, we propose our approach and toolset named CONFINE, which we developed with the intent of enabling process mining on process event data from multiple providers while preserving the confidentiality and integrity of the original records. To ensure that the presented interaction protocol steps are secure and that the processed information is hidden from both involved and external actors, our approach is based on a decentralized architecture and consists of trusted applications running in Trusted Execution Environments (TEE). In this demo paper, we provide an overview of the core components and functionalities as well as the specific details of its application.
\end{abstract}

\maketitle

%% Tool metadata table: if a metatadum is not applicable for your tool 
%% (e.g., source code repository for a commercial application), write 
%% N/A in the appropriate field
%% Please do not move/resize this table - as it should stay after the 
%% abstract, before the first section - nor change the labels in the 
%% metadata description column
\begin{table*}[h]
\footnotesize
\begin{tabular}{ll}
\toprule
Metadata description               & Value \\
\midrule
Tool name                          & CONFINE \\
Current version                    & 1.0 \\
Legal code license                 & Apache 2.0 \\
Languages, tools and services used & Go, Python \\
Supported operating environment    & GNU/Linux  \\
Download/Demo URL                  & \href{https://github.com/Process-in-Chains/CONFINE.git}{\nolinkurl{github.com/Process-in-Chains/CONFINE.git}}  \\
Documentation URL                  & \href{https://github.com/Process-in-Chains/CONFINE/blob/main/README.md}{\nolinkurl{github.com/Process-in-Chains/CONFINE/blob/main/README.md}} \\
Source code repository             & \href{https://github.com/Process-in-Chains/CONFINE}{\nolinkurl{github.com/Process-in-Chains/CONFINE}} \\
Screencast video                   & \href{https://youtu.be/Oaoo6gS_4tw?si=hY0ztcbUrGO8PjIr}{\nolinkurl{youtu.be/Oaoo6gS_4tw?si=hY0ztcbUrGO8PjIr}} \\
\bottomrule
\end{tabular}
\end{table*}

\section{Introduction}
\label{sec:intro}

Collaboration between multiple organizations is essential in today's highly competitive and development-oriented business environment. %Sharing common goals enables organizations to reduce costs and improve efficiency and performance. 
By aligning around shared goals, organizations can streamline operations, reduce redundancies, and ultimately improve efficiency, performance, and cost-effectiveness. In this context, inter-organizational process mining enables the coordination of joint efforts, improves overall transparency, and allows for benchmarking~\cite{van2011intra}. Nevertheless, despite the numerous benefits, a number of potential issues may arise, primarily related to data confidentiality. Information is an asset, and even in a collaborative environment companies are unwilling to share sensitive data required to execute process mining algorithms with their partners~\cite{liu2009challenges}. Allowing sensitive operational data to move across organizational boundaries inevitably raises issues related to data privacy and security, potentially exposing the data to unauthorized access and misuse~\cite{muller2021trust}.
For instance, we may consider a hospitalization process that involves three different parties: a \Actor{Hospital}, a \Actor{Pharmaceutical company}, and a \Actor{Specialized clinic}. In this scenario, two other entities wish to uncover information on the inter-organizational process for reporting and auditing purposes: the National Institute of Statistics of the country where the three organizations reside and the University that hosts the hospital~\cite{Jans.Hosseinpour/IJAIS2019:ActiveLearningProcessMiningForAuditing}. The \Actor{Hospital}, \Actor{Specialized clinic}, and \Actor{Pharmaceutical company} have a partial view of the overall dynamics of the inter-organizational process as they record operations pertaining to their own parts. As a result, each player stories a separate event log partition. It would be mutually beneficial for all parties involved to have access to the findings of an aggregate data analysis, integrating the event log partitions %with the process data 
yielded by the collaborating organizations. Nevertheless, % there is 
an intrinsic divergence of interests emerges: %between the various parties involved, 
%  which precludes the possibility of 
granting access to traces to other organizations may reveal information the parties do not want to disclose.
The ability to conduct process mining on data from multiple sources must, therefore, meet the necessity to safeguard the privacy of the entities involved and to guarantee the confidentiality of the information, ensuring that the local event log is never given away completely in-clear. 

To solve this conundrum, we present CONFINE, our recently developed approach and toolkit designed to enhance collaborative information system architectures with secrecy-preserving process mining capabilities. To secure information secrecy during the exchange and elaboration of data, our solution resorts to \emph{Trusted Execution Environments} (TEEs)~\cite{DBLP:conf/trustcom/SabtAB15}. These are hardware-secured contexts, which guarantee code integrity and data confidentiality before, during, and after their utilization. Owing to these characteristics, CONFINE lets information be securely transferred beyond an organization's perimeter. Therefore, computing nodes other than the information provisioners can aggregate and elaborate the original, unaltered process data in a secure, externally inaccessible vault. Also, CONFINE is capable of providing these guarantees while demanding low computation overhead and providing scalability.

\begin{figure}[tb]
 \centering
	\includegraphics[width=.9\textwidth]{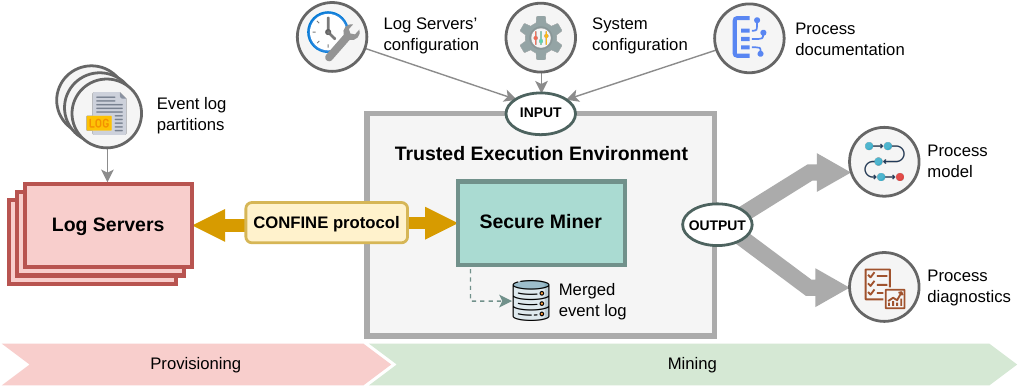}
	\caption{An overview of the CONFINE architecture}
	\label{fig:confine_scheme}
\end{figure}

\section{The CONFINE Framework}
In the following, we present the core concepts of CONFINE by introducing the main components and discussing how they interact in the CONFINE protocol.
\label{sec:confine}
\paragraph{High-level architecture.}
\label{CONFINE:architecture}
\Cref{fig:confine_scheme} displays a high-level schematization of the CONFINE framework.
Our architecture involves different information systems running on multiple machines. An organization can take at least one of the following roles, depending on the tasks they take on:
\begin{inparadesc}
\item[provisioning] if it delivers local partitions of event logs to be collaboratively mined (e.g., the hospital in our example);
\item[mining] if it applies process mining algorithms using event logs retrieved from provisioners (e.g., the National Institute of Statistics).
\end{inparadesc}
In our solution, every organization hosts one or more nodes, incorporating components that depend on the roles played. 
%
% An organization assumes the provisioning role when its objective within the cooperation context is to share data related to its processes. When assigned this role, the 
The nodes hosted by provisioning organizations implement the \Compo{Log Server} component. 
% An organization takes on the miner role when it aims to perform mining operations on the shared processes. In this case, the node implements the 
The data provided by \Compo{Log Server}s is fed into the \Compo{Secure Miner} components, which are deployed on nodes hosted by mining organizations.
Notice that every \Compo{Secure Miner} retrieves process data from one to many \Compo{Log Server}s, as each of the latter can offer different partitions of the event log (for example, the \Actor{Hospital} records activities that are not visible to the \Actor{pharmaceutical company}, and vice-versa). In CONFINE, multiple organizations can perform mining tasks by hosting one or more \Compo{Secure Miner}s, eliminating the need to depend on a central authority. This reflects the decentralized nature of our approach.

%Nodes come endowed with a \Compo{Log Server} or a \Compo{Secure Miner} component, or both. 
CONFINE allows miners to request data from provisioners that are part of the inter-organizational context to perform mining operations while ensuring the secrecy and confidentiality of the event logs exchanged. These properties are ensured since the \Compo{Secure Miner} is a trusted application executed within a trusted execution environment (TEE). %\todo[inline]{piccole info tee?\\ By enforcing memory access restrictions, TEE aims to prevent one application from reading or altering the memory space of another, thus enhancing system security.}
TEEs create isolated contexts separate from the operating system, safeguarding code and data with hardware-based encryption mechanisms. They use dedicated CPU instructions to manage encrypted data within a reserved memory portion~\citep{costan2016intel}. The CPU encrypts this memory with a random decryption key generated at each power cycle. By enforcing strict memory access controls, TEEs prevent applications from accessing or altering each other's memory space, thus enhancing system security.
The CONFINE protocol securely interconnects \Compo{Log Server}s with the \Compo{Secure Miner}s. 

\paragraph{The CONFINE protocol.}
\label{CONFINE:protocol}
% When a \Compo{Secure Miner} wants to perform mining operations on the shared process, the CONFINE protocol is triggered. 
In our protocol, the interaction between components is divided into four sequential steps, namely
\begin{inparaenum}[\itshape(i)\upshape]
 \item \emph{initialization}, 
 \item \emph{remote attestation}, 
 \item \emph{data transmission}, and 
 \item \emph{computation}.
\end{inparaenum}
The \Compo{Secure Miner} starts the protocol execution with the \textit{initialization} step, during which it gets informed about the case distribution of the log partitions in the \Compo{Log Server}s; this data request includes identity evidence of the mining organization. 
In this phase, e.g., the \Compo{National Institute of Statistics} requests the \Actor{Hospital}'s \Compo{Log Server} to list aggregate information about the cases they can share in order to pre-plan the subsequent data requests.
Next, the \textit{remote attestation} phase occurs. The purpose of remote attestation is to establish trust between the \Compo{Log Server}s and the \Compo{Secure Miner}. % The identification that takes place in the previous step is not enough because it is only based on attributes sent by the miner. 
Since the actual process data are going to be transmitted next, a stronger certification is required.
This phase is thus based on the RATS RFC standard~\cite{rfc9334} (the basis of TEE attestation schemes) and has three main objectives: 
\begin{inparaenum}[\itshape(i)\upshape] \item to provide the \Compo{Log Server}s evidence that the data request for a log partition originates from a trusted application running within a TEE; 
\item to verify that the trusted application is indeed the authentic \Compo{Secure Miner} software entity; 
\item to identify the owner of the \Compo{Secure Miner} application.
\end{inparaenum}
Once the trusted nature of the \Compo{Secure Miner} is verified, %the \Compo{Log Server}s can proceed to transmit their case partitions in the \textit{data transmission} phase. % Since in the previous phase the \Compo{Secure Miner} has collected the list of cases that each \Compo{Log Server} has. It requests the cases that are present in all the \Compo{Log Server}s in order to provide for reconstructing the complete case by performing a merge operation. 
it retrieves from the \Compo{Log Server}s their log partitions to internally merge the cases.
Based on the information retrieved during the initialization phase, it plans the requests accordingly.
Each \Compo{Log Server} retrieves its event log and filters it based on the case id specified by the miner. Given the typically limited capacity of a TEE working memory, \Compo{Log Server}s are required to divide the filtered log into separate segments of up to a certain size. In the \textit{computation} phase, the process mining technique selected by the organization is applied to the received data. The mining organizations %performing the data mining 
can use the \Compo{Secure Miner} in either an \emph{incremental} or a \emph{non-incremental} manner. The distinction between these approaches concerns the timing in which the computation phase is executed. With the incremental approach, the \Compo{Secure Miner} begins the computation phase concurrently with the data transmission phase: As soon as all \Compo{Log Server}s have sent their log segments pertaining to a specific case, %, allowing the \Compo{Secure Miner} to fully merge a case without missing any data, 
the mining algorithm is executed to produce and update a partial mining result. % on a case-by-case basis. 
In contrast, with the non-incremental approach, the \Compo{Secure Miner} waits until all % requested cases have arrived at the \Compo{Log Server}s. Once all the data is received, 
full partitions are received from the \Compo{Log Server}s, and only then it starts the mining algorithm.
% \Compo{Secure Miner} performs the final merge operation, after which the mining operation is conducted.
The incremental approach tends to save on TEE memory (the full log does not need to get loaded before mining begins), but it poses a restriction on the mining algorithm in use as it must be apt for treating partial inputs and produce updates.

\section{Maturity}
\label{sec:maturity}
\begin{figure}[tb]
 \centering
	\includegraphics[width=.9\textwidth]{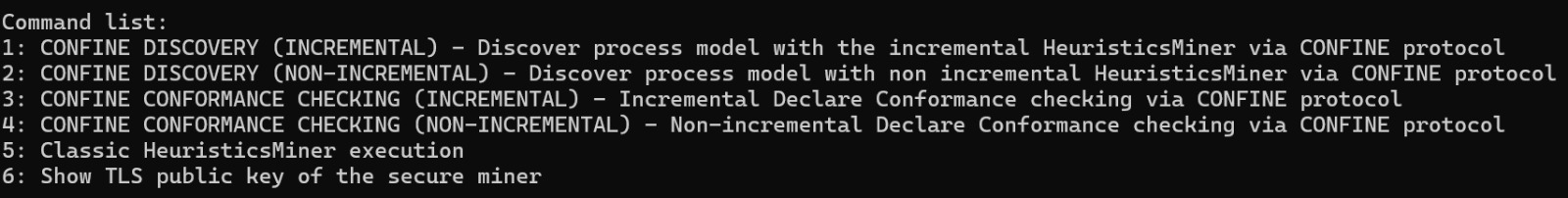}
	\caption{Screnshot of the \Compo{Secure Miner}'s terminal interface}
	\label{fig:confineInterface}
\end{figure}

To showcase the capabilities of CONFINE, we propose a prototype implementation and run it with artificially generated and real-world data.
% of the \Compo{Secure Miner} alongside the \Compo{Log Server}. 
Our \Compo{Secure Miner} implementation is an Intel-SGX%
\footnote{\url{https://www.intel.com/content/www/us/en/developer/tools/software-guard-extensions/overview.html}. Accessed:~September~26,~2024.} 
trusted app developed in GO.%
%\footnote{\url{https://go.dev}. Accessed: \today }
We adopt the EGo%
\footnote{\url{https://www.edgeless.systems/products/ego}. Accessed:~September~26,~2024.} 
framework to deploy the \Compo{Secure Miner} trusted app into the Intel SGX TEE. As depicted in \cref{fig:confine_scheme}, the \Compo{Secure Miner} takes as input a configuration of \Compo{Log Server}s, and a set of parameters to execute the protocol, and process documentation (a collective term indicating cues for the mining algorithms, such as a collaborative process specification for conformance checking). Users can interact with \Compo{Secure Miner} using a terminal interface through which commands are forwarded to the TEE. \Cref{fig:confineInterface} shows a screenshot of this interface. We developed the \Compo{Log Server} in GO. Communication between the \Compo{Secure Miner} and \Compo{Log Server} relies upon the HTTP protocol secured via TLS.

Our \Compo{Secure Miner} implementation incorporates the \textit{Heuristics Miner}~\cite{weijters2006process} algorithm for process discovery and a porting of the PM4Py \textit{Conformance Declare}%
\footnote{\url{https://processintelligence.solutions/static/api/2.7.11/pm4py.algo.conformance.declare.html}. Accessed:~September~26,~2024.} 
algorithm for the conformance checking of declarative process specifications~\cite{di2022declarative}. The results produced by the HeuristicsMiner can be examined using workflow net visualization tools like \emph{WoPeD}~\cite{freytag2017woped}. %\footnote{\url{https://woped.dhbw-karlsruhe.de}. Accessed: \today.} 
We provide both an \textit{incremental} variant, which updates partial results as complete cases are collected, and a \textit{non-incremental} variant, starting only after the entire merged event log is collected. We validated our prototypical implementation of CONFINE on real-world event logs, including BPIC 2013~\cite{bpic2013} and Sepsis~\cite{seps}, as well as a synthetic log generated from a healthcare scenario. We evaluated the convergence of the CONFINE protocol and the scalability of the \Compo{Secure Miner}'s runtime memory usage. These tests were conducted in both simulation mode and SGX native mode.

We plan to improve the CONFINE toolset in different directions. %At present, the \Compo{Secure Miner} implementation is capable of handling only one configuration of \Compo{Log Server}. We intend to make the \Compo{Secure Miner} able to manage multiple clusters of \Compo{Log Server}s, gaining insights from different process settings. \todo{Too brainy. Truncated.}
Given the stringency of the hardware requirement imposed by CONFINE, we envision real-world production scenarios involving the \Compo{Secure Miner} instances being deployed on remote machines owned by service providers. To take steps towards this direction, we plan to implement remote APIs through which mining organizations can remotely submit commands to the \Compo{Secure Miner}s and securely receive the output of the computation. At present, the \Compo{Secure Miner} prototype offers a limited choice of process mining algorithms. We aim to widen the spectrum of applicable algorithms by designing a plug-in integration system enabling implementers to add computation modules that are auditable by \Compo{Log Server} via \textit{remote attestation}. Furthermore, the integration of a graphical user interface would facilitate the interaction with the \Compo{Log Server} and the \Compo{Secure Miner} trusted app. These improvements pave the path for future work.

%\todo[inline]{
%Esplicitare lo stato prototipale dell'implementazione\\
%log mondo reale BPIC2013 E SEPSIS\\
%test intel non solo simulazione\\
%Future work: Secure Miner fornito da service provider in cui ogn'uno puo create la propria instanza--> API client per controller il secure miner remoto\\
%Future work: Integrazione di piu algoritmi di PM--> questo implica che ci siano piu measurement (cambiare il termine) del secure miner da dover condividere con i log servers\\
%}

\section{Availability}
A more detailed description of the CONFINE approach is available in~\cite{goretti2024trusted}. Our implementation can be downloaded from \href{https://github.com/Process-in-Chains/CONFINE}{\nolinkurl{github.com/Process-in-Chains/CONFINE}}.
The \texttt{readme} file of the repository guides the installation of the \Compo{Secure Miner} and the \Compo{Log Server} components. The video demonstration of CONFINE is included in the repository's \texttt{readme} and can be directly watched at \href{https://www.youtube.com/watch?v=Oaoo6gS\_4tw}{\nolinkurl{www.youtube.com/watch?v=Oaoo6gS\_4tw}}.
\begin{acknowledgments}
This research work was partly funded by MUR under PRIN grant B87G22000450001 (PINPOINT), the Latium Region under PO~FSE+ grant B83C22004050009 (PPMPP), Sapienza University of Rome under grant RG123188B3F7414A (ASGARD), and by the EU-NGEU under the NRRP MUR grant PE00000014 (SERICS).
\end{acknowledgments}

%%
%% Define the bibliography file to be used
\bibliography{bibliography}

\appendix

%\section{Getting a license}
%If the tool requires a license, this appendix should describe how to obtain a (temporary) license. The procedure to obtain the license must not disclose the identity of the reviewers. This appendix will not be included in the final version for the proceedings, if the demo is accepted.

\end{document}